\documentstyle[11pt,newpasp,twoside]{article}
\begin{document}
\title{Analysis of Balmer Profiles of early type stars}
\author{Mercedes Moll\'a, Angeles I. D\'{\i}az, Mar Alvarez Alvarez} 
\affil{Departamento de F\'{\i}sica Te\'{o}rica, Universidad Aut\'{o}noma
 de Madrid, Campus de Cantoblanco, 28049 Spain}
\author{Rosa M. Gonz\'{a}lez Delgado}
\affil{Instituto de Astrof\'{\i}sica de Andaluc\'{\i}a, C.S.I.C., Apdo. 3004, 
18080 Spain}


\section{Analyzing Profiles to Discriminate among Early Stellar Types}

  We have analyzed the high resolution profiles presented by
Gonz\'{a}lez Delgado \& Leitherer (1999) for the
Balmer and He lines, given for solar metallicity, 4000 K $\leq \rm
Teff \leq $50000 K, and $0 \leq\log{g} \leq 5 $, and we have studied
the potential use of these profiles to discriminate between different
type of early stars, which dominate the spectrum of star-forming
regions. Here we present our analysis for the H8 line. Details for the
whole set of Balmer and He lines will be given elsewhere.

 From this analysis we conclude that profiles are deeper for decreasing
Teff and constant $\log{g}$, while keeping the same shape. On the
contrary, for increasing $\log{g}$ but with a same Teff, they show a
{\sl widening} of the wings, and a depth almost constant. This way,
stars having almost identical EW values, may show very different
profiles.  We use their characteristics to separate the stellar
parameter effects, gravity and effective temperature, by using two
simultaneous methods:

1. We use 2 indices to characterize each profile: a) Index $ D$ in
magnitudes, which does not depend on gravity but depends on Teff; and
b) the core equivalent width $\rm EW_{c}$, defined as the area limited
by $\lambda_{c}-3$ and $\lambda_{c}+3$ ($\lambda_{c}$ is the central
wavelength) for each feature, which shows a clear dependence on
gravity, due to the widening of the wings, and an almost independence
of Teff. With a diagnostic diagram of $\rm EW_{c}$ {\sl vs.} $ D$ it
is possible to separate both effects, gravity and effective
temperature.
 
2. We have constructed growth curves by computing the equivalent width
EW(j) for each $\Delta \lambda (j)=\lambda_{c}\pm 0.3 \times j$.
These growth curves EW(j) increase smoothly up to the total values EW,
which are reached at $\lambda$'s more or less close to $\lambda_{c}$,
according to the concentration of each profile.  As profiles result
deeper when Teff decreases, but their shape is the same for constant
${g}$, the growth curves, normalized to the total EW, result similar
for all Teff. The growth curves for a same Teff and different
$\log{g}$ are almost equal in the core, but EW(j) increases with
increasing gravity as the profiles become wider.

\end{document}